\documentclass[3p,times,twocolumn]{elsarticle}

\usepackage{ecrc}

\volume{00}

\firstpage{1}

\journalname{Nuclear Physics B Proceedings Supplement}

\runauth{T.K. Gaisser for IceCube}

\jid{nuphbp}

\jnltitlelogo{Nuclear Physics B Proceedings Supplement}

\usepackage{amssymb}

\usepackage{lineno}

\usepackage[figuresright]{rotating}

\begin{document}

\begin{frontmatter}




\title{Primary spectrum and composition with IceCube/IceTop}


\author{Thomas K. Gaisser for the IceCube Collaboration}

\address{Bartol Research Institute and Department of Physics and Astronomy\\
University of Delaware\\
Newark, DE 19716 USA}

\begin{abstract}
IceCube, with its surface array IceTop, detects three different components 
of extensive air showers: the total signal at the surface, 
GeV muons in the periphery of the showers and TeV muons in the deep array of IceCube.  
The spectrum is measured with high resolution from the knee to the ankle with IceTop.  
Composition and spectrum are extracted from events seen in coincidence 
by the surface array and the deep array of IceCube.  
The muon lateral distribution at the surface is obtained from the data 
and used to provide a measurement of the muon density at 600 meters from the shower core 
up to 30 PeV.  Results are compared to measurements from other experiments to obtain 
an overview of the spectrum and composition over an extended range of energy.  
Consistency of the surface muon measurements with hadronic interaction models 
and with measurements at higher energy is discussed. 
\end{abstract}


\end{frontmatter}

\section{Introduction}
\label{sec:introduction}

\begin{figure}[thb]
\includegraphics[width=1.1\columnwidth]{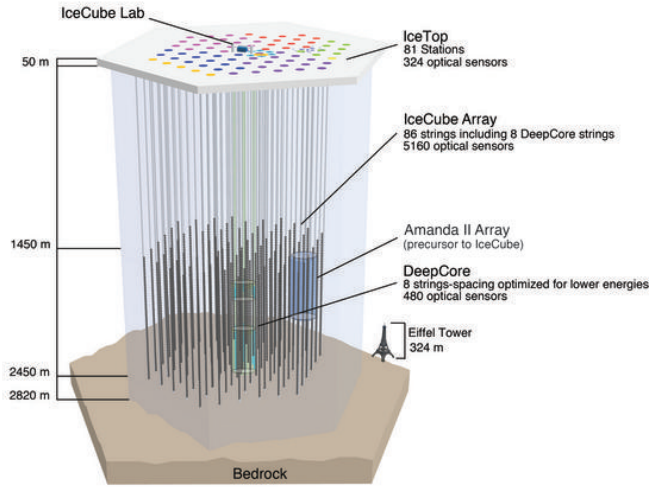}
\caption{Layout of the IceCube Observatory at the South Pole. 
\label{fig:IceCube}
}
\end{figure}

\begin{figure}[thb]
\centerline{
\includegraphics[width=1.0\columnwidth]{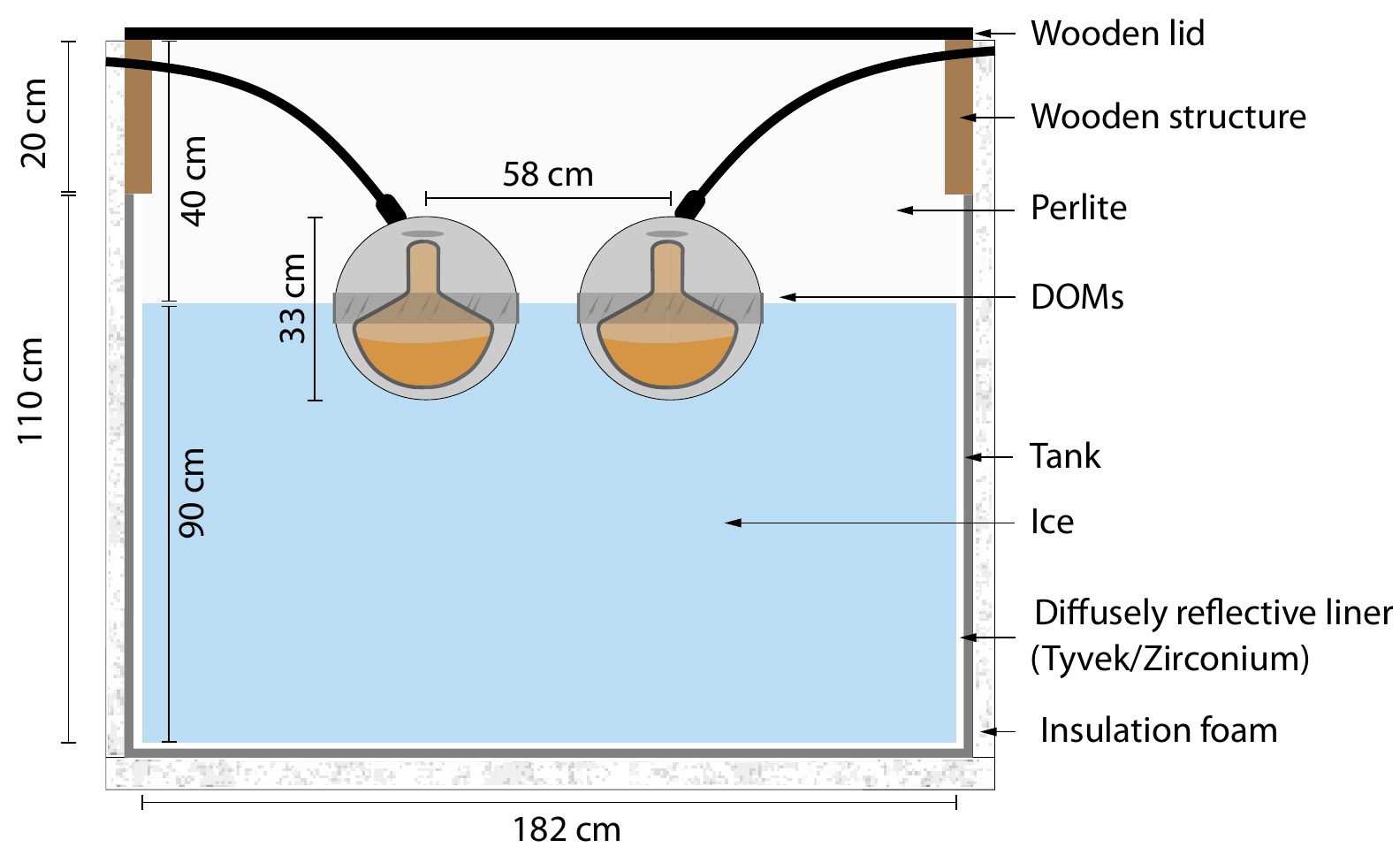}
}
\caption{Cross section of an IceTop tank~\cite{IceCube:2012nn}.
\label{fig:tank}
}
\end{figure}

The IceCube Neutrino Observatory includes a surface detector above the
deep array as illustrated in Fig.~\ref{fig:IceCube}.  With an area of $\approx1$~km$^2$,
IceTop is sensitive to the primary spectrum from PeV to EeV.  The surface array
consists of 81 stations each with two tanks separated from each other by $10$~m
and filled with clear ice~\cite{IceCube:2012nn}.  Each tank (see Fig.~\ref{fig:tank}) 
is viewed by two digital optical modules (DOMs),
one running at high gain and the other at low gain to achieve a dynamic range of $\approx\!10^4$
for the energy deposited in each tank.  The IceTop DOMs are fully integrated into
the data acquisition system of IceCube so that timing across the full array is
accurate to $\approx3$~ns.  The spectrum measurement with IceTop
benefits from the high altitude of the array (equivalent to a depth of $\approx690$~g/cm$^2$),
which allows a measurement of the spectrum with very good energy resolution.

Events with trajectories that pass through IceTop and the deep array of IceCube
can be reconstructed in both parts of the detector.  The signal in the deep array
is due to energy deposition by muons sufficient energy at production
to reach the deep array (500 GeV minimum and typically $\sim$TeV).
The ratio of 
muons in the deep detector to the shower size measured at the surface is sensitive to primary
composition.  This measurement is related to the classic $\mu/e$ ratio measured at the
surface.  Heavy nuclei of a given primary energy produce more muons than protons
of the same energy in both cases.  However, the systematics of the two analyses are
different because the TeV muons are from higher energy interactions in the shower than
the GeV muons at the surface.  Making both measurements on the same set of showers
therefore has the potential to improve the understanding of systematic differences in hadronic interaction
models.  We return to this point in Section~\ref{sec:surface-muons} below.

\begin{figure}[htb]
\centerline{
\includegraphics[width=1.0\columnwidth]{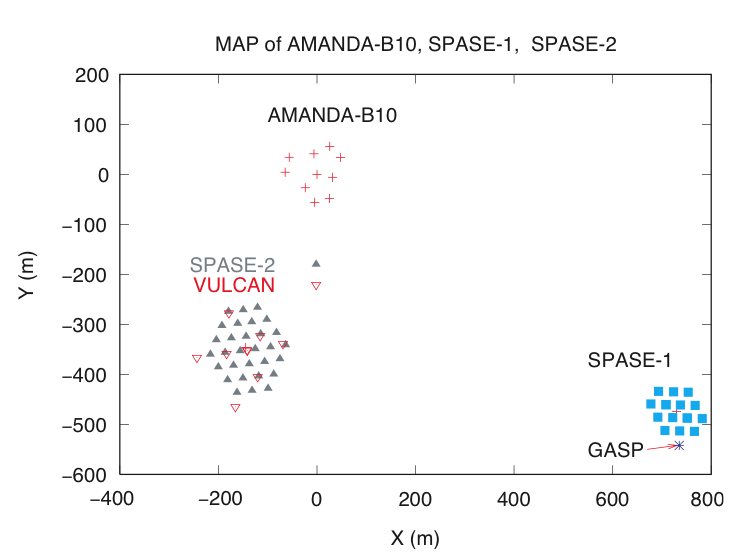}
}
\caption{Configuration for the muon survey of AMANDA-B10 by SPASE~\cite{Ahrens:2004dg}. 
The SPASE-1 array was grid NW of the South Pole dome on the station side of
the airplane skiway.  SPASE-2 and AMANDA were in the ``dark sector" grid west of
the skiway.  SPASE-2 included a sub-array of atmospheric Cherenkov light detectors called
VULCAN (inverted red triangles).
\label{fig:SPASE-AMANDA}
}
\end{figure}

The first deep underground muon detector near Cornell University
in upstate New York~\cite{Barrett:1952xyz} also set up a small air shower array on the surface.  
With underground detectors of order $1$~m$^2$ at a depth of $600$~m
and surface detectors spaced by $\approx 60$~m,
the aperture was tiny ($\sim 0.01$~m$^2$sr).  The first serious measurement of coincidences
between a surface array above a deep detector was EASTOP-MACRO~\cite{Bellotti:1989pa}. 
The aperture for observing coincident events was $\sim 100$~m$^2$sr, and the
muon energy threshold at the surface was $\approx1.3$~TeV.  
The South Pole Air Shower Experiment (SPASE-2)~\cite{Dickinson:2000hr} was used in coincidence
 with the Antarctic Muon and Neutrino Array (AMANDA), the forerunner of IceCube, 
for a composition analysis with coincident events~\cite{Ahrens:2004nn}.  Its aperture
 was also $\sim 100$~m$^2$sr.
 For comparison, the aperture of IceCube for coincident
 events is $\approx 0.25$~km$^2$sr.
 The earlier air-shower experiment, SPASE-1, was decommissioned in 1995, 
 but it also ran in coincidence
 with AMANDA during construction.  The configuration of a two-dimensional muon
 survey of AMANDA-B10 from the surface arrays~\cite{Ahrens:2004dg} is shown in 
 Fig.~\ref{fig:SPASE-AMANDA}.  

\section{Energy spectrum using IceTop only}
\label{sec:IceTop only spectrum}

Showers in IceTop are reconstructed by fitting a lateral distribution function
to the observed signal, taking account of arrival time fluctuations as described
in Ref.~\cite{IceCube:2012nn}.  Because snow accumulates at a different rate
over each IceTop tank, measured signals are corrected before fitting to the
lateral distribution function.  The correction is made with a simple exponential
absorption factor, $\exp{X_i/\lambda\cos\theta}$, with the snow depth 
interpolated between biennial measurements at each tank.  
The average spacing between stations in IceTop is $125$~m.  Correspondingly, the
shower size is characterized by the fitted signal ($S_{\!125}$) at $125$~m
perpendicular from the shower trajectory.  The energy spectrum is obtained
from comparison of the measured size spectrum to a Monte Carlo simulation
of shower size vs primary energy for different groups of nuclei.  

\begin{figure}[!thb]
\centerline{
\includegraphics[width=1.0\columnwidth]{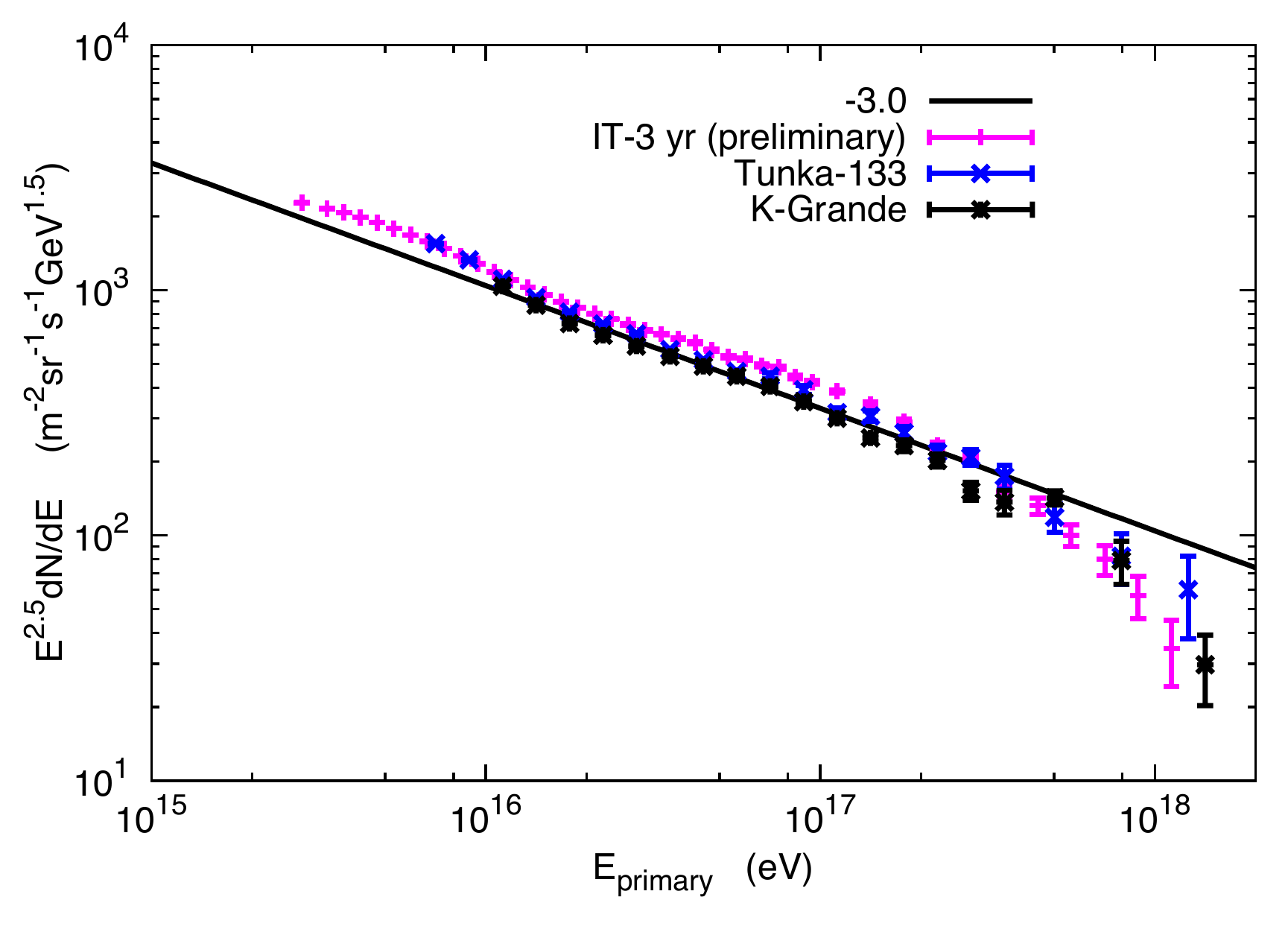}
}
\caption{Energy spectrum from the knee to the EeV from three years of IceTop~\cite{Aartsen:2015awa5}
compared to a pure power law (solid line) and to data from KASCADE-Grande~\cite{Apel:2012rm}
and TUNKA~\cite{Prosin:2014x1}.
\label{fig:all-part-compare}
}
\end{figure}

One of the main systematic uncertainties is the composition model (relative
contribution of different mass groups vs energy) assumed to obtain the
mean primary energy for a given $S_{\!125}$.  The H4a model of Ref.~\cite{Gaisser:2012zz} 
is used to make the size to energy conversion.  The sensitivity to composition
is checked~\cite{Aartsen:2013wda} by making the conversion at each of four
zenith angle bins assuming pure protons and assuming pure iron.  Under the
assumption of pure protons, the spectrum at the larger zenith angle is lower
than that for the vertical.  Under the assumption of pure iron for the primary
composition, the angular dependence of the spectra in the angular bins is
reversed.  This behavior reflects the fact that proton shower penetrate
more deeply than iron showers for a given primary energy.  
When the H4a model of composition is used,
the spectra obtained at the four different zenith angles are closer to each
other.  In principle, the composition could be inferred by adjusting
the relative fractions at each energy to get the same primary spectrum
for each zenith bin, as required by the fact that the true spectrum
is independent of the zenith angle at which it is measured.  In practice,
such an approach is difficult because of fluctuations.  The angular dependence
of the spectra reconstructed assuming H4a is used as a measure of the
systematic uncertainty from composition.

The energy spectrum measured in 2010-11 with the nearly complete IceTop detector (IT-73 with
73 of 81 stations in operation)~\cite{Aartsen:2013wda} showed clearly that the spectrum between the
knee and the ankle cannot be described by a single power law.  The high resolution
measurement with IceTop clarifies the structure seen in previous experiments.  The
same analysis has now been applied to three years of IceTop data (2010-2013), with
the data from the complete 81 station array analyzed using only the IT-73 tanks for consistency
with the first year analysis~\cite{Aartsen:2015awa5}.  The three-year analysis includes an improved treatment
of the time-dependent correction for
snow above the detector.  By comparing reconstructed
events in areas with deeper snow to those in areas with little or no
snow, the effective attenuation parameter was optimized to 2.1~m for the first
year of the analysis and 2.25~m for the subsequent two years.  (For snow
density of $\approx 0.4$~g/cm$^2$, 2.1~m corresponds to an effective
attenuation length of 84~g/cm$^2$.)  Work is underway to account for
the fact that it is mainly the electromagnetic component of the signal
that is affected by snow~\cite{Aartsen:2015awa9}.

\begin{figure}[!htb]
\centerline{
\includegraphics[width=1.0\columnwidth]{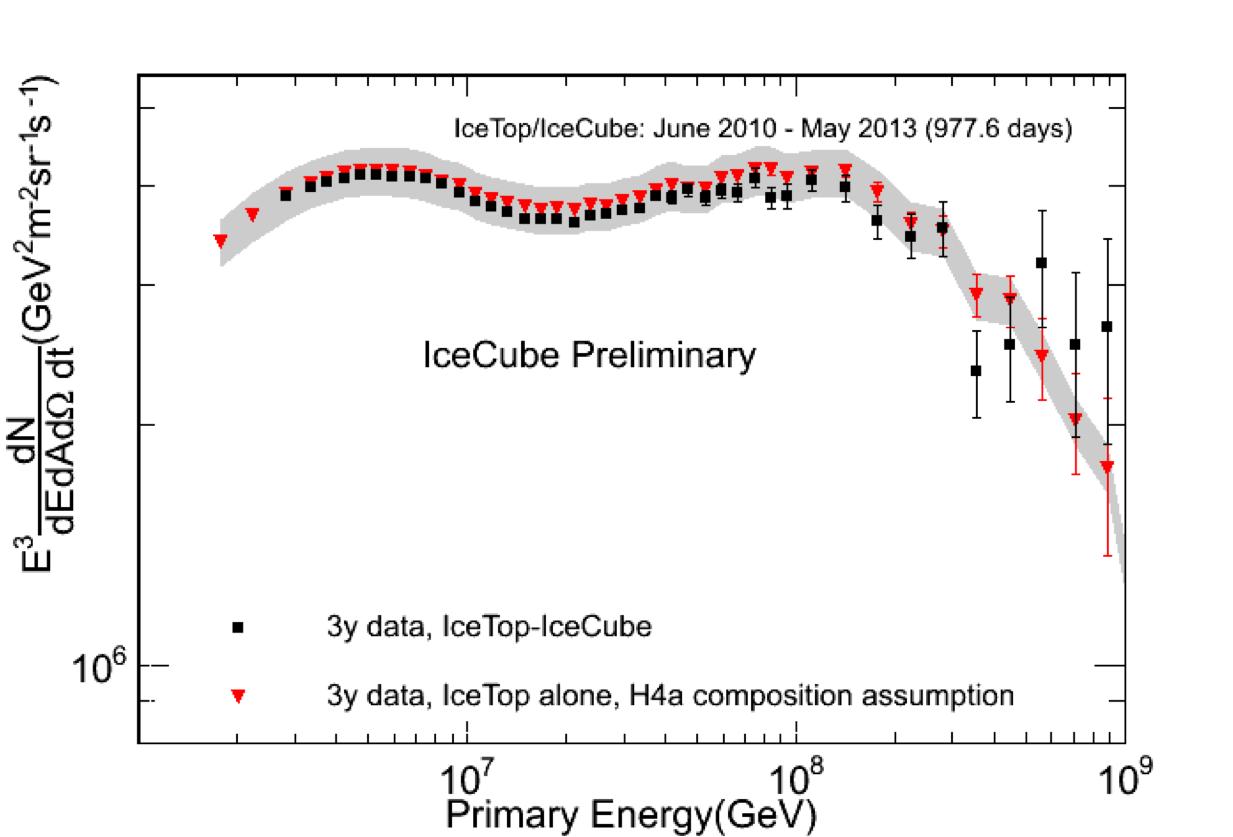}
}
\caption{Comparison of the IceTop only spectrum (red points) with the spectrum from 
the IceCube coincident analysis (black points).
\label{fig:spectrum2}
}
\end{figure}

The IceTop 3-year data are shown
along with data from KASCADE-Grande~\cite{Apel:2012rm} and TUNKA~\cite{Prosin:2014x1} 
in Fig.~\ref{fig:all-part-compare} compared to an $E^{-3}$ differential power law
shown by the solid line.  There is a hardening of the spectrum around $2\times 10^{16}$~eV
and a steepening above $2\times 10^{17}$~eV, sometimes referred to as the second knee.

\section{Coincident events}
\label{sec:coincident-events}

With a sample of coincident events, each of which is measured both 
by IceTop and by the deep array of IceCube,
the degeneracy between energy and composition with the IceTop only analysis 
can be removed.  The coincident event analysis~\cite{Aartsen:2013lla} uses
a neural network (NN) to determine both energy and composition from the three-year
sample of coincident events that are well contained and reconstructed in
both IceTop and IceCube.  An updated description of the coincident analysis is
given in Ref.~\cite{Aartsen:2015awa5}.  Figure~\ref{fig:spectrum2} compares
the energy spectrum obtained from the coincident analysis with the IceTop only
spectrum.  The good agreement below $10^8$~GeV confirms that the composition systematic has
been dealt with in a reasonable way in the IceTop only analysis, which has
the higher statistics.

The principal observables on which the network is trained (using simulated data) are $S_{\!125}$,
$\cos\theta_{\rm zenith}$ and $E_\mu^{1500}$, the reconstructed energy loss per meter of muons 
in the shower as it enters the deep array at 1500 m. 
The reconstruction is based on the observed energy losses within the detector.
In addition, two measures of the number of
stochastic energy losses in the reconstructed in-ice track
(moderate and high) are used.  The NN is trained and tested on half the showers simulated
with Sibyll-2.1~\cite{Ahn:2009wx} and FLUKA~\cite{Battistoni:2007zzb} 
for protons, helium, oxygen and iron primaries.  The output variables
are the shower energy and a measure of $\langle\ln(A)\rangle$ for each event.
Applying the trained NN to the other half of the simulated data leads
for each energy bin to a set of histograms for each mass group.
Events in these ``template" histrograms are classified by a proxy for $\ln(A)$.
Applying the NN to the data leads to an energy estimate for each event
and a single histogram for each energy bin.
The $\langle\ln(A)\rangle$ for each bin is obtained by finding the best fit of the
four template histograms to the data histogram for the corresponding energy bin.

\begin{figure}[thb]
\centerline{
\includegraphics[width=1.0\columnwidth]{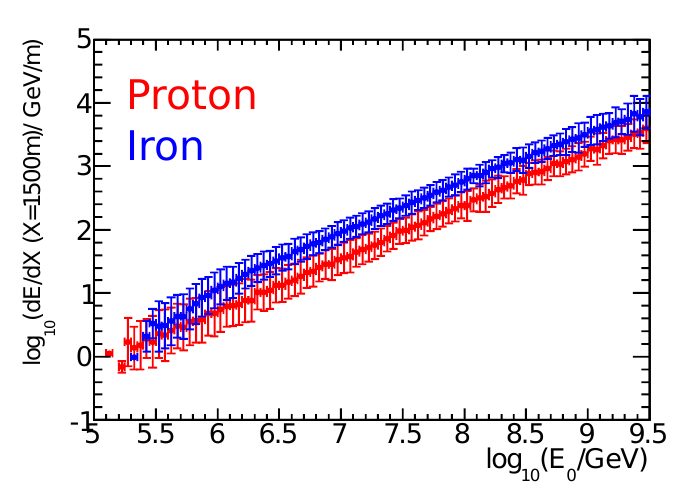}
}
\caption{Reconstructed energy loss as a function of 
primary energy for
showers initiated by protons (red) and by iron nuclei (blue).
\label{fig:Eloss}
}
\end{figure}

\begin{figure}[thb]
\centerline{
\includegraphics[width=1.0\columnwidth]{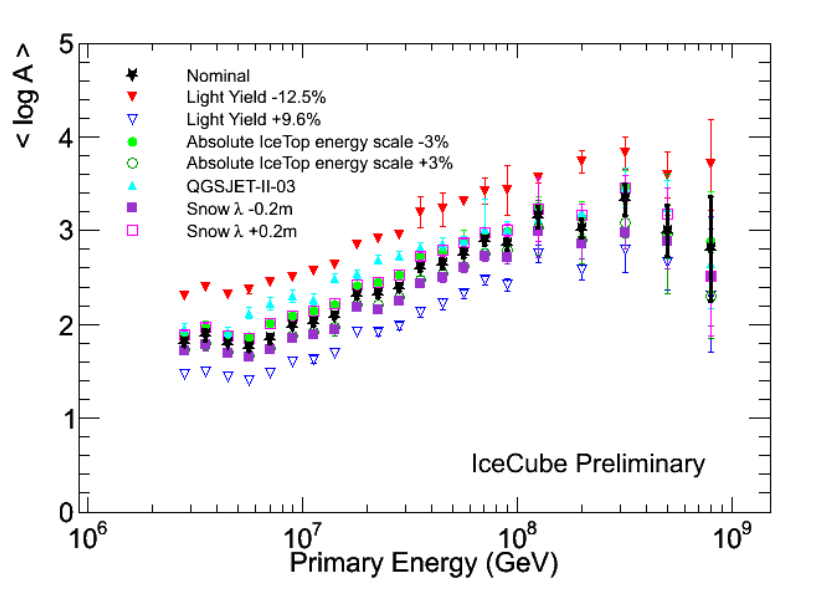}
}
\caption{Energy dependence of $\langle\ln(A)\rangle$ from the coincident analysis
for various assumptions on each of the sources of systematic uncertainty.
\label{fig:logA}
}
\end{figure}

The main composition-dependent variable in the NN analysis is $E_\mu^{1500}$.
Its sensitivity is illustrated in Fig.~\ref{fig:Eloss} from simulations
of protons and iron.  An important systematic uncertainty in the coincident analysis
is absolute calibration of the light yield in the detector.  To the extent that
the main source of differences among interaction models is the number of
$\sim$TeV muons, those systematic uncertainties will scale similarly to
the light yield.  Figure~\ref{fig:logA} shows $\langle\ln(A)\rangle$ from the
coincident analysis at its nominal value (black stars) and scaled according to the 
various systematic effects listed.

\begin{figure}[thb]
\centerline{
\includegraphics[width=1.0\columnwidth]{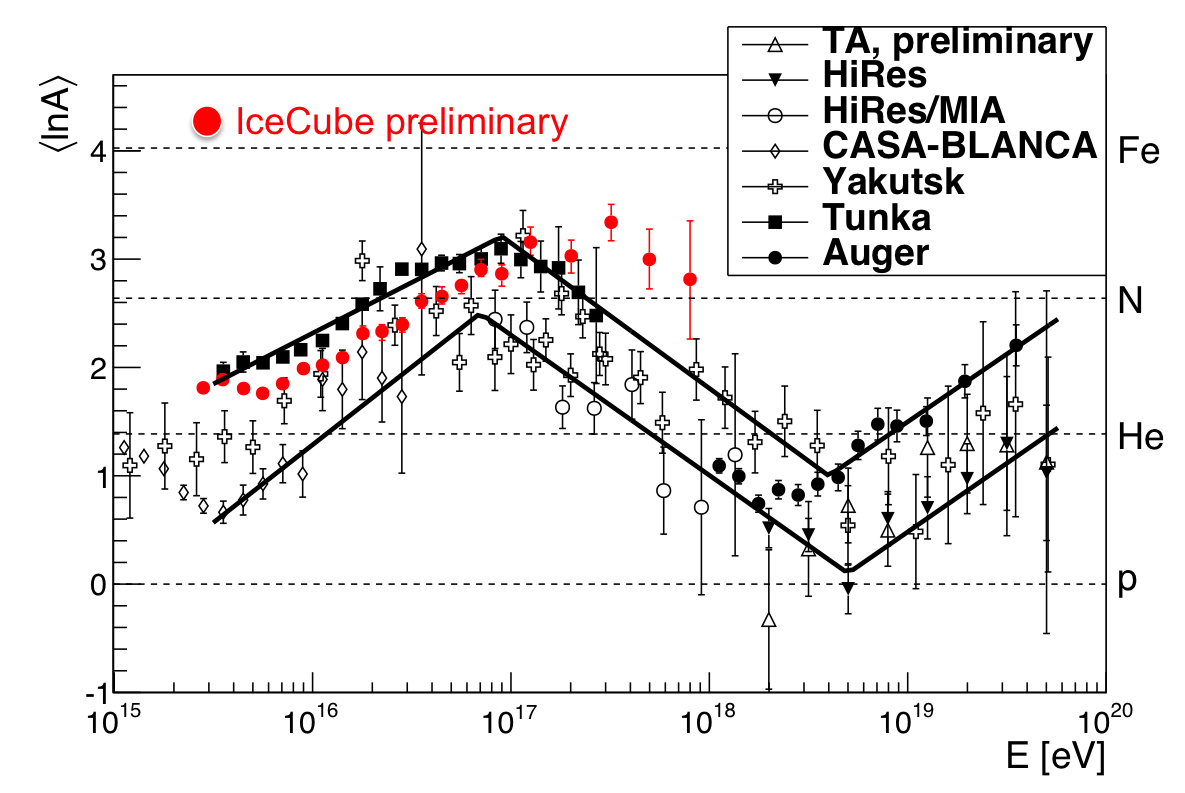}
}
\caption{Comparison of $\langle \ln(A)\rangle$ from the IceCube coincident analysis
with a compilation of data~\cite{Kampert:2012mx}.  (See text for discussion.)
\label{fig:logAcompare}
}
\end{figure}

The central values from the IceCube coincident analysis 
are shown as the red points in Fig.~\ref{fig:logAcompare} superimposed on
the compilation from the review paper of Kampert and Unger~\cite{Kampert:2012mx}.
The values of $\langle\ln(A)\rangle$ are obtained in Ref.~\cite{Kampert:2012mx} by interpolating 
measured values of shower maximum between values of $X_{\rm max}$ for protons and
iron from simulations.  Here we show the diagram interpreted with Sibyll-2.1 to be
consistent with the coincident analysis.  The solid lines are included in the figure
from Ref.~\cite{Kampert:2012mx} to indicate the range of the data summarized.

\section{Surface muons}
\label{sec:surface-muons}

The DOMs in IceTop tanks record waveforms from the Cherenkov light produced 
by charged particles  with speeds above the Cherenkov threshold in the clear ice.
The amount of light depends on the track length in the tank but not on the identity of
the particle(s) that produced it.  There are, however, several possibilities
for obtaining some information about the muon content of air showers with IceTop.
For example, there is potential information in the structure of the waveforms which
might serve to distinguish muons from electromagnetic signals, which are 
primarily due to conversion of photons in the tanks.

\begin{figure}[!htb]
\centerline{
\includegraphics[width=1.0\columnwidth]{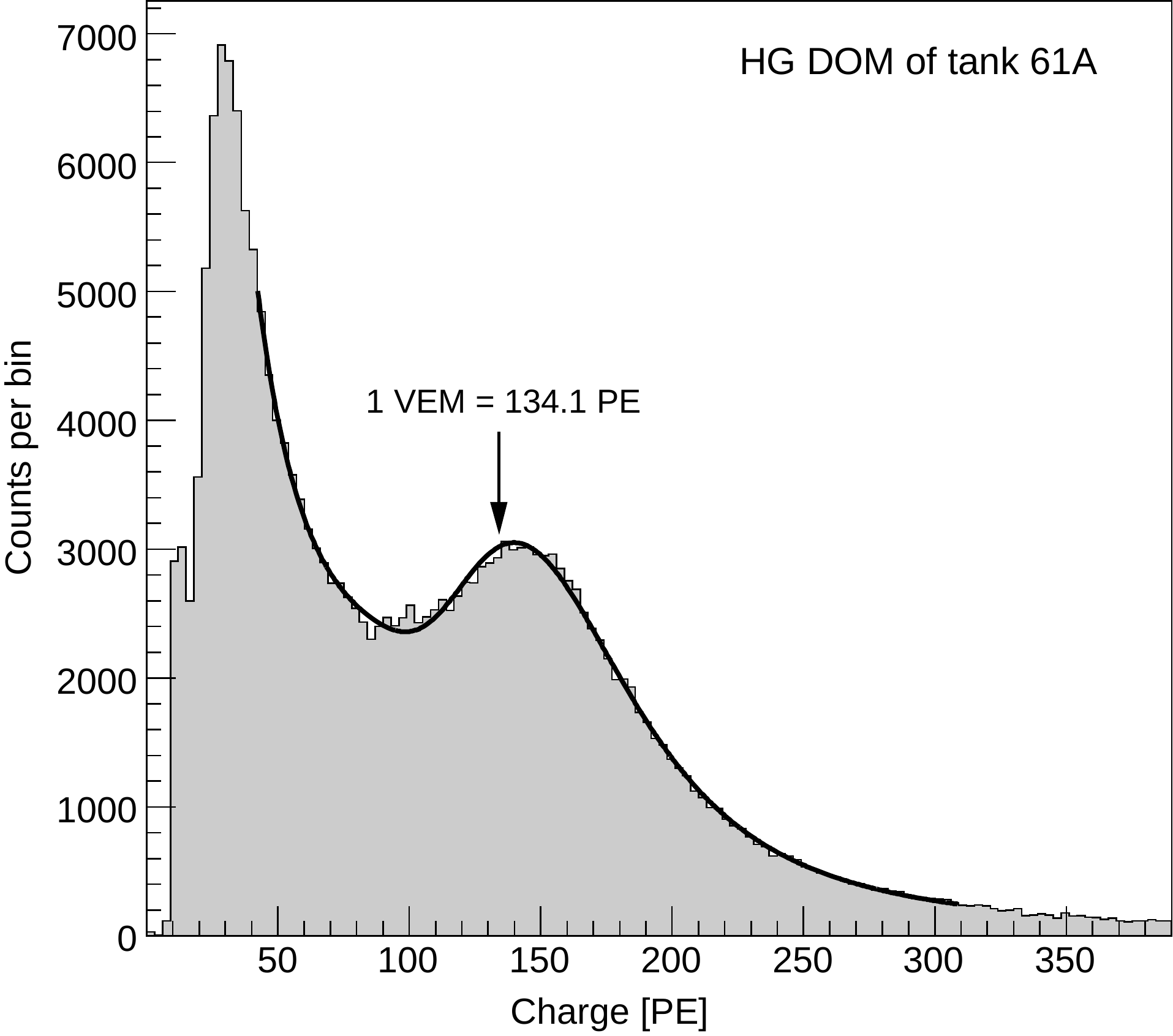}
}
\caption{Distribution of signals for a calibration run for the high-gain DOM
in tank 61A.  The definition of one vertical equivalent muon (VEM) is indicated.
(From~\cite{IceCube:2012nn}.)
\label{fig:muon-calib}
}
\end{figure}

\begin{figure}[!htb]
\centerline{
\includegraphics[width=1.0\columnwidth]{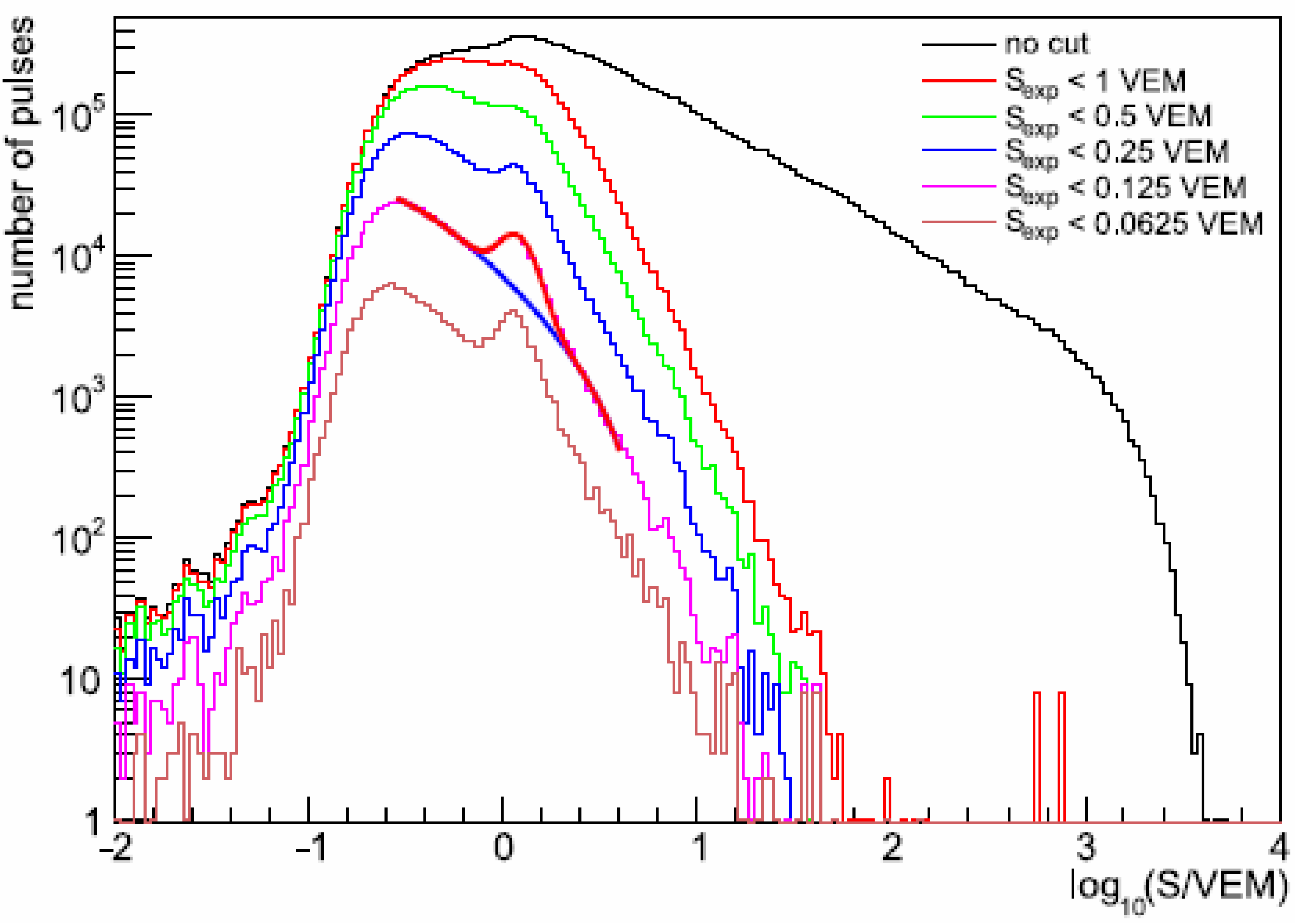}
}
\caption{Distribution of tank signals in many showers classified by successively
larger cuts in the core distance defined in terms of the expected signal.
(From~\cite{IceCube:2012nn}.)
\label{fig:lateral}
}
\end{figure}

\begin{figure}[!htb]
\centerline{
\includegraphics[width=1.0\columnwidth]{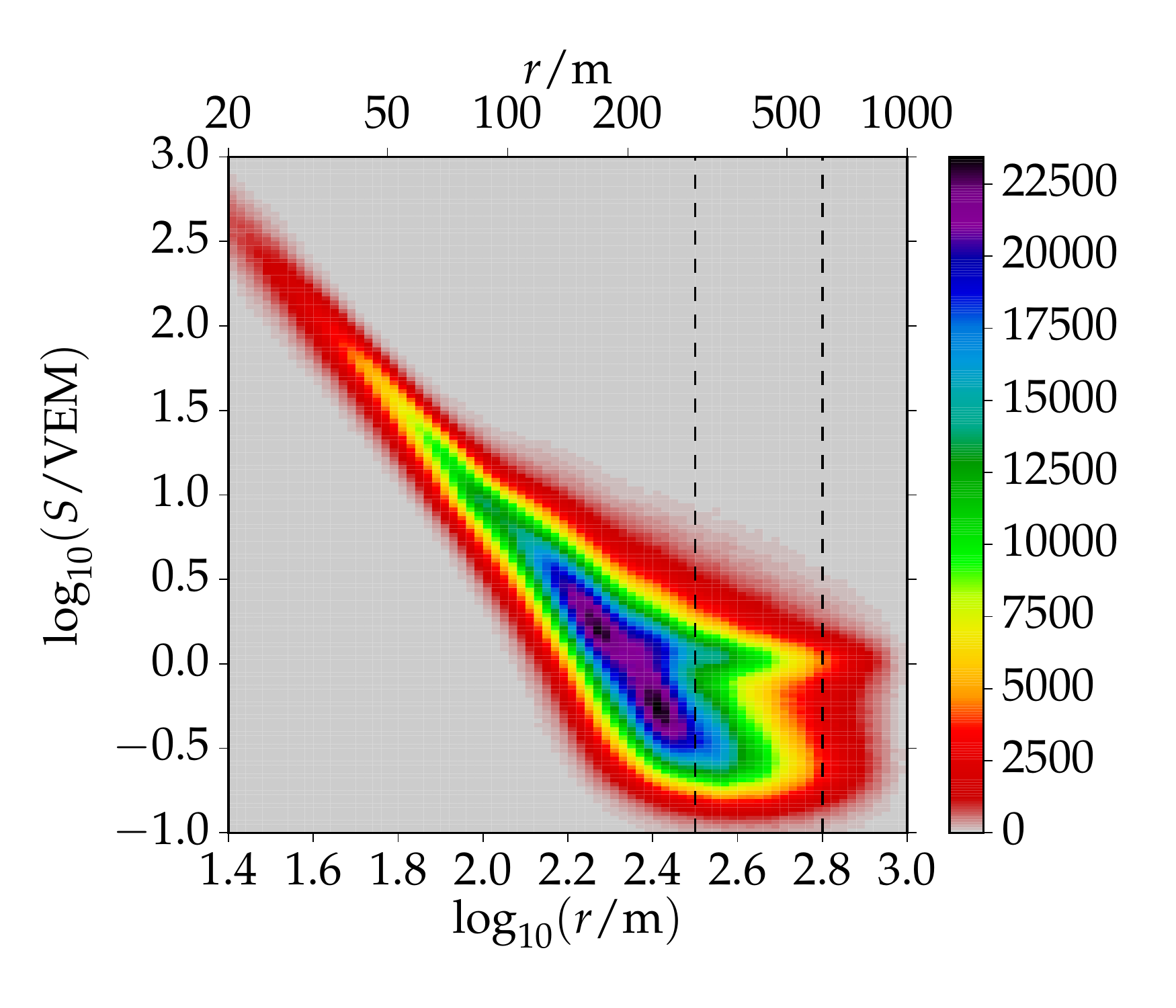}
}
\caption{Two-dimensional distribution of signals in showers with
energies $\approx3$~PeV in the zenith angle
bin around $13^\circ$ as a function
of the VEM signal and core-distance.
\label{fig:thumb}
}
\end{figure}


A simpler method is to make use of the fact that the characteristic charge
distribution of muons passing through a tank is understood well
from the calibration procedure.  Figure~\ref{fig:muon-calib} shows the
charge distribution from one calibration run for a high-gain DOM in one tank. 
The calibration data are obtained from uncorrelated hits collected without
an air shower trigger, so they are from the continuous flux of photons, electrons
and muons produced by interactions of relatively low energy cosmic rays in
the atmosphere.

Signals in IceTop tanks are defined in units of VEM obtained 
by monthly calibration runs for each tank (see Fig.~\ref{fig:muon-calib}).  In particular, the lateral distribution
of an air shower is expressed in terms of VEM as a function of core-distance.  In
the inner region of a shower, a signal of $\sim1$~VEM can be produced either by
a combination of electromagnetic quanta with appropriate total track length or
by a muon or by a combination of the two (if the muon stops in the tank).  In the
outer region of the shower, however, a signal near one VEM is likely to be
from a muon.  The "outer region" is defined as the distance beyond which
the fitted lateral distribution for a shower falls below one VEM.  Figure~\ref{fig:lateral}
illustrates how the muon peak becomes more pronounced at larger core distance. 
Figure~\ref{fig:thumb} is a two dimensional representation of the same 
information.  The "thumb" at 1 VEM reflects the muons.

\begin{figure}[!htb]
\centerline{
\includegraphics[width=1.0\columnwidth]{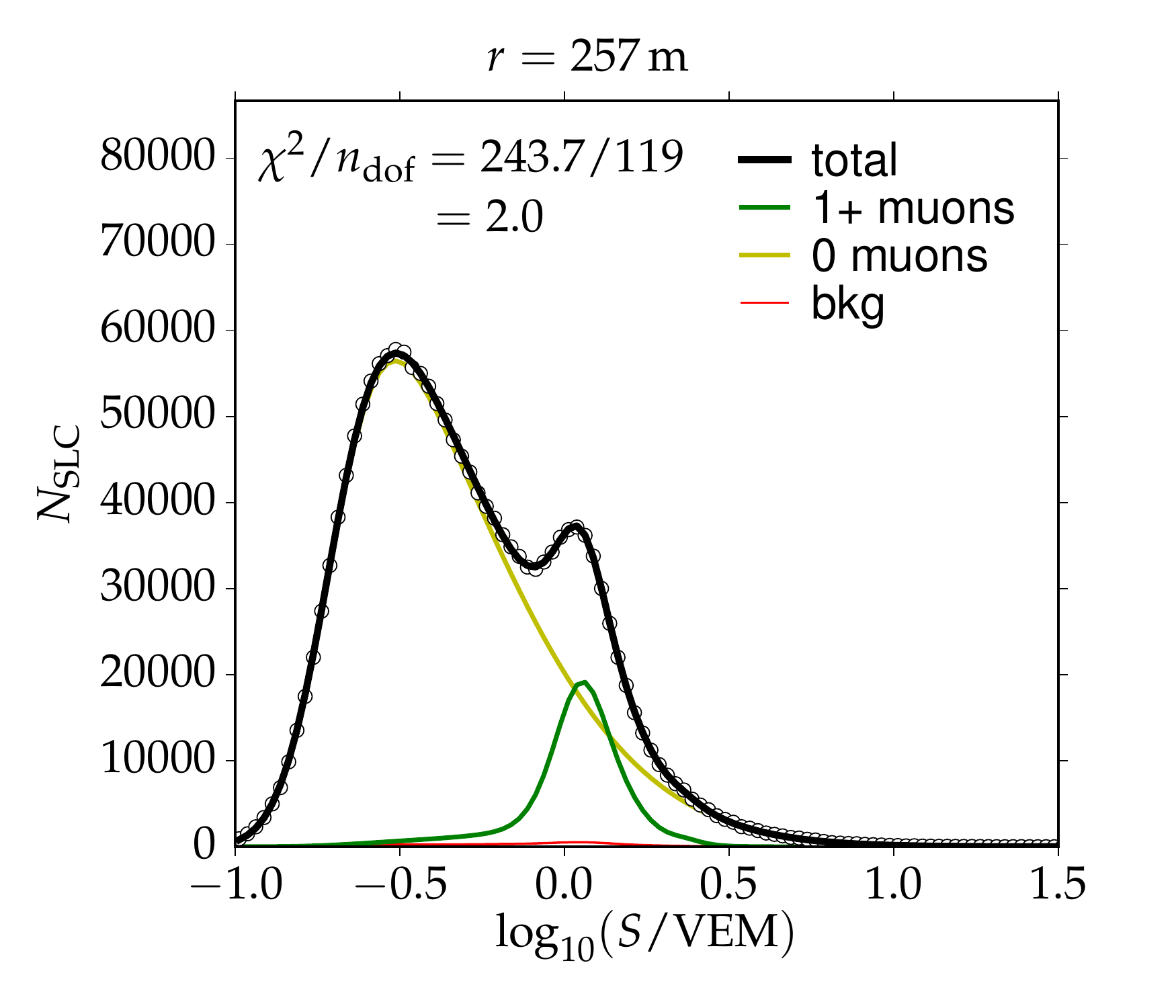}
}
\caption{Data in tanks in the radial bin around $257$~m corresponding to Fig.~\ref{fig:thumb}.
(Figure from~\cite{Aartsen:2015awa3}.).
\label{fig:lat1}
}
\end{figure}

\begin{figure}[!htb]
\centerline{
\includegraphics[width=1.0\columnwidth]{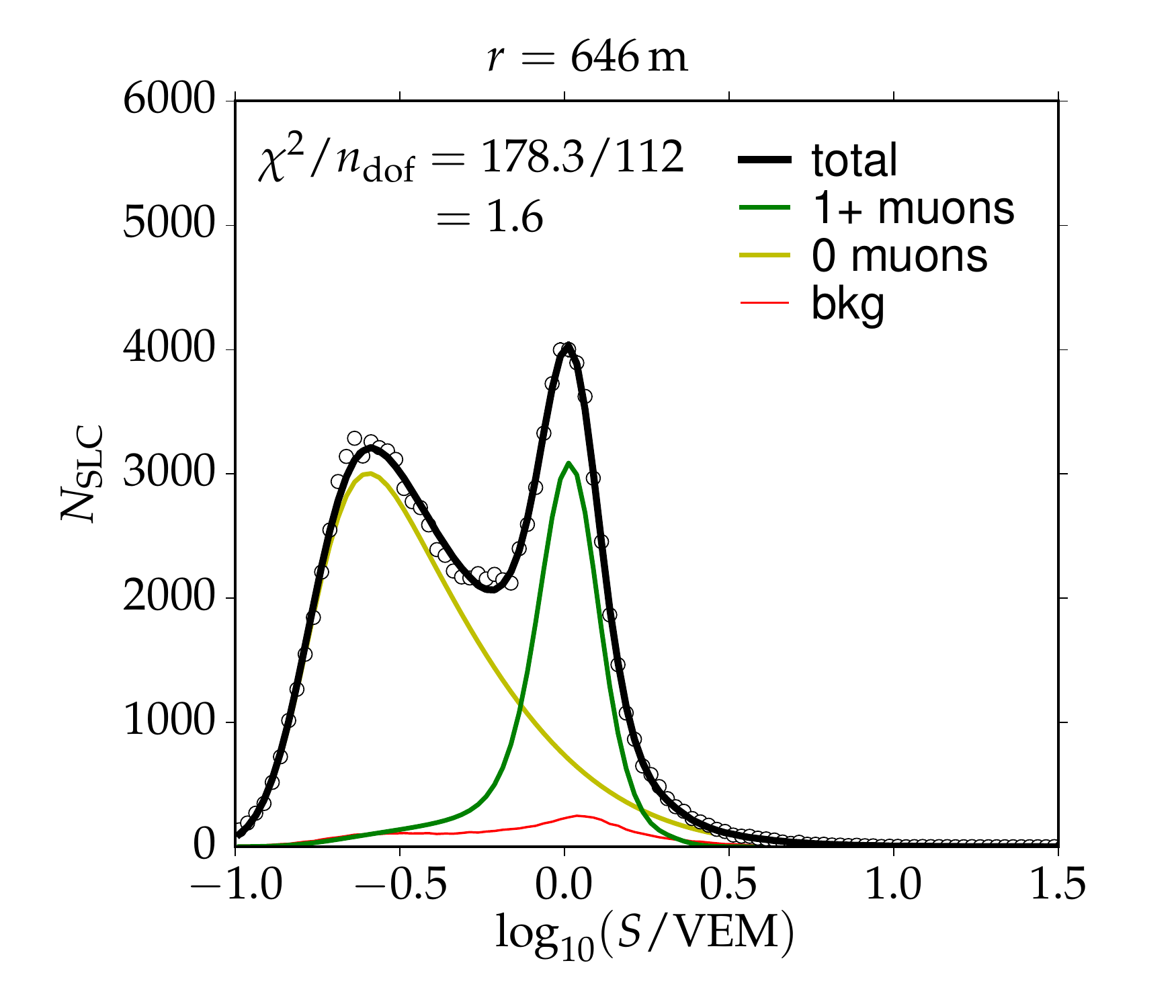}
}
\caption{Data in tanks in the radial bin around $646$~m corresponding to Fig.~\ref{fig:thumb}.
(Figure from~\cite{Aartsen:2015awa3}.)
\label{fig:lat2}
}
\end{figure}

Muon signals in air showers cannot, however, be fit directly from the shape of
the muon peak in calibration runs because the air shower context is different.
In addition, the showers need to be analyzed as a function of zenith angle and energy.  
The analysis starts by making
distributions like that in Fig.~\ref{fig:thumb} for each bin of zenith angle and energy.
Then the data are further divided into bins of core distance, defined as distance
in the shower plane perpendicular to the reconstructed trajectory of the shower.
Two examples are shown in Figs.~\ref{fig:lat1} and~\ref{fig:lat2} at distances
corresponding to the vertical dashed lines in Fig.~\ref{fig:thumb}.  These figures
show the data sample for an energy bin around $3$~PeV centered around core distances of $257$ and $646$~m.
The muon peak becomes increasingly prominent relative the the electromagnetic component 
as distance increases.

The data are fit by three components as indicated in the two examples: 
(1) one or more muons, (2) 0 muons (electromagnetic), and (3) background.
(The small background of accidental hits not related to the shower is
determined from the distribution of hits outside the time windows of events in the sample.)
The shape for the muon contribution is obtained starting with GEANT4 simulations of the tank response
to one muon as a function of zenith angle.  Because a tank can be hit by more than one
muon, the actual muon signal distribution may be broader than for a single muon.  
The shape of the $\ge 1\,\mu$ peak
 is characterized
by $\langle N_\mu\rangle$ in which the shapes for 1, 2, and 3 muons are combined with weights
according to a Poisson distribution determined by fitting with $\langle N_\mu\rangle$
as a parameter.  The relative normalization of the
sum of the electromagnetic only (major) and background (small) contribution to the fit
must correspond to the total Poisson probability of having 0 muons,
while the normalization of the muon contribution (1) is
the Poisson probability of having at least one muon. 
The detailed procedure is described in Ref.~\cite{Gonzalez:2015dda}.  
Once the mean muon number in a given radial bin is fixed, the
muon density is obtained by dividing by the total projected area
of tanks in that radial bin.
The lateral distribution of muons found in this way can be described by the Greisen
function,
\begin{equation}
\rho_\mu(r)=\rho_\mu(r_0)\left(\frac{r}{r_0}\right)^{-3/4}\left(\frac{r_1+r}{r_1+r_0}\right)^{-\gamma}
\label{eq:greisen}
\end{equation}
where $r_1=320$~m and the reference radius is $r_0=600$~m.  The normalization
parameter, $\rho_\mu(r_0)$ and the slope parameter $\gamma\approx 2.5$ are
fitted for each primary energy, where the relation between $S_{\!125}$ and primary
energy is determined as in Ref.~\cite{Aartsen:2013wda}.

\begin{figure}[!htb]
\centerline{
\includegraphics[width=1.0\columnwidth]{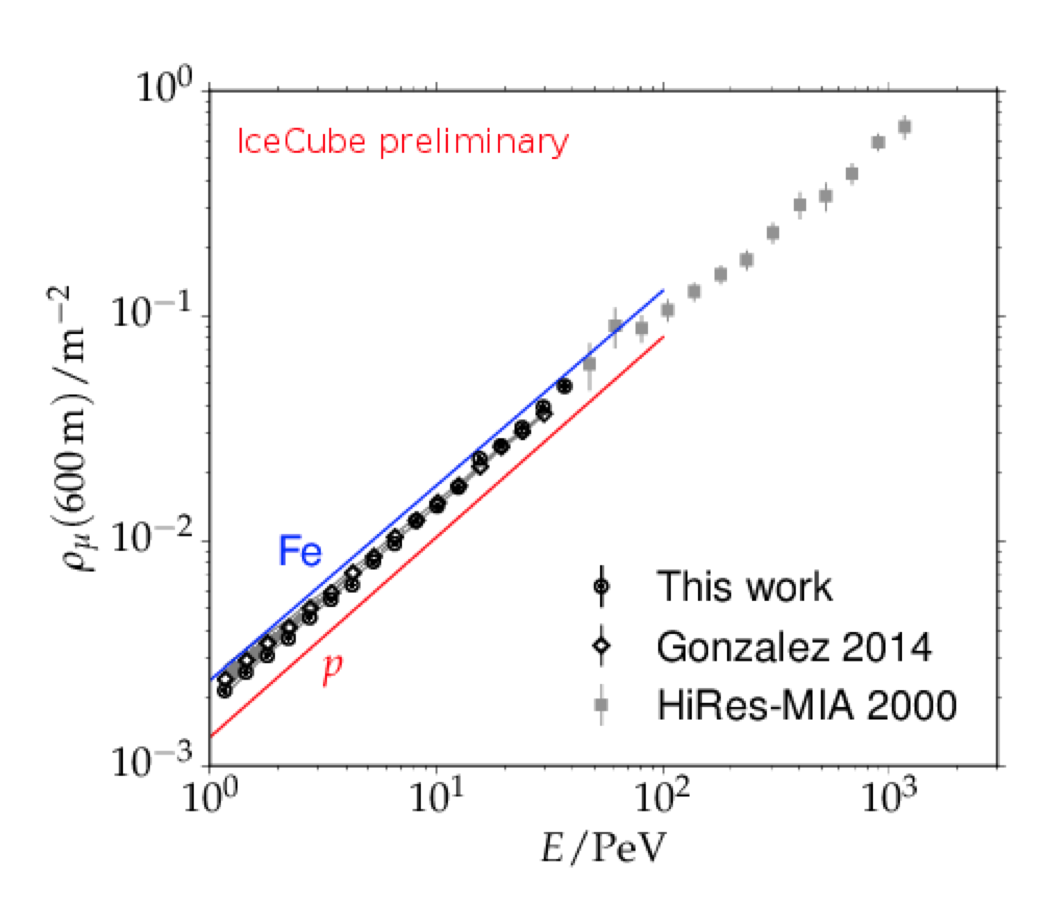}
}
\caption{Muon density at $600$~m as a function
of energy.  (Figure from~\cite{Aartsen:2015awa3}.)
\label{fig:rho-muon}
}
\end{figure}

The result is a set of muon lateral distributions determined directly
from IceTop data for a range of zenith angles and primary energies.
From these lateral distributions, the muon density at $600$~m is
determined as a function of energy and compared in Fig.~\ref{fig:rho-muon} 
to data at higher energy from Hi-Res-MIA~\cite{AbuZayyad:1999xa} in
Fig.~\ref{fig:rho-muon}.  The densities expected from primary iron (blue, upper line)
and from protons (red, lower line) are also shown (derived from SIBYLL 2.1).

The method was first presented and preliminary results shown in 
Ref.~\cite{Gonzalez:2015dda}.
The later analysis~\cite{Aartsen:2015awa3} presented at ICRC 2015 includes a
comparison with simulations (with Sibyll 2.1), shown here in Fig.~\ref{fig:rho-muon}.
The result is consistent with expectations
to the extent that the measurements are between protons and iron.
In particular, there is no evidence for a significant excess of
muons in data up to 30 PeV compared to simulations, 
in contrast with the situation at 10 EeV where there appear to be
more muons in data than expected~\cite{Aab:2014pza}.  Also, the post-LHC
models seem to have 30\% more muons than Sibyll 2.1 
at least up to 30~PeV (J. Gonzalez, private communication).

The muon content of showers at the surface is sensitive to primary composition.
As for the coincident events, more muons are expected for events generated by
heavy primaries than by light primary nuclei of the same energy.  However,
the muons come from different stages of shower development and reflect
different properties of the parent hadronic interactions in the two cases. 
Muons at the surface typically have energies of a few GeV and are produced
by decay of mesons produced in interactions of order $100$~GeV.  In contrast,
the $\sim$~TeV muons in the coincident analysis should be produced earlier
in the shower and descended from hadronic interactions an order of magnitude higher.
For this reason, a full composition analysis with surface muons will be
important for comparison with the coincident analysis.  Differences in hadronic
interaction models are likely to affect the two analyses differently.
Finding consistency between the two may therefore be helpful in placing
constraints on interaction models as well as on composition.

Building on the tools developed for the measurement of muons at the surface
in IceTop, it is possible to develop an analysis that will return
the muon content on an event-by-event basis.  This is done by
fitting each shower with two lateral distribution functions,
one for the electromagnetic component and the other for muons.
In this analysis, individual signals are assigned a probability
of containing a muon based on the known properties of the
muon and electromagnetic signals in IceTop discussed above.  
The concept and preliminary examples were presented at the 2015
ICRC~\cite{Aartsen:2015awa6}.  Because of the sensitivity of the
muon number to primary composition, this would make it possible to
assign a probability of light vs. heavy primary to each event.
In addition to the value for composition analysis, such a method
would also make possible a composition-dependent version
of the IceCube cosmic-ray anisotropy analysis~\cite{Aartsen:2012ma}.
It would also allow a better snow correction by identifying the
electromagnetic contribution.

\section{Other approaches}

There are several other approaches to using muons in IceCube to
help determine the primary cosmic-ray composition.  A comparison
of muon bundles in the deep part of IceCube to simulations shows
promise because of its large reach in energy, from tens of TeV to
well into the EeV energy range~\cite{Aartsen:2015nss}.  The analysis
uses the connection between primary mass and number of muons and finds
a steadily increasing mass, consistent with the coincident event
analysis of Section~\ref{sec:coincident-events} and somewhat in tension 
with the summary of Ref.~\cite{Kampert:2012mx}
above $10^{17}$~eV (compare Fig.~\ref{fig:logAcompare}).

Still other approaches involve what might be called the geometry
of muons in air showers.  The typical size of a muon bundle in
the deep detector below $1.5$ km.w.e. is less than the string spacing of 125~m.
It is therefore possible to identify individual outlying muons separated
from the main bundle by more than the string spacing.  The lateral 
distribution of $\sim$TeV muons in IceCube is discussed in Ref.~\cite{Abbasi:2012kza}.
To achieve a seperation of, for example $135$~m, a vertical muon with sufficient
energy to reach IceCube ($500$~GeV) needs a transverse momentum of $6$~GeV/c
if it is produced at an altitude of $25$~km, the typical interaction
height for a heavy nucleus~\cite{Aartsen:2015awa2}.  The relation between 
heavy nuclei and protons for production of high transverse momentum
particles is complex.  On the one hand, heavy nuclei have first interactions
higher in the atmosphere, but on the other the energy per nucleon is lower,
so the fraction of high-transverse momentum is lower.  The analysis
therefore depends both on simulations of the detector response and on
the hadronic interaction models used.  Finding a consistent interpretation
thus has the potential of clarifying both aspects.

Another approach under investigation is to use the timing of
muons at large distances to reconstruct the distribution of
muon production heights as in Auger~\cite{Aab:2014dua}.  The goal is
to measure the muon production profile and hence to obtain the muon
depth of shower maximum as a composition-dependent parameter.  This
analysis should naturally be associated with the measurement of surface muons
discussed in the previous section.

\section{Future}

Motivated by the observation of high-energy astrophysical 
neutrinos in IceCube~\cite{Aartsen:2013jdh,Aartsen:2014gkd},
planning for an expanded detector is underway~\cite{Aartsen:2014njl}.
The basic concept~\cite{::2015dwa} is to increase the neutrino detection volume by
an order of magnitude using 120 new strings with $\approx 240$~m spacing around
the present detector, which has a string spacing of $125$~m.  Studies of
ice properties with the present detector show that the vertical instrumentation
can be increased by 27\% (1360-2621~m compared to 1450-2450~m) at present.

The primary goal of IceCube Gen2 is to obtain sufficient statistics to
characterize the astrophysical spectrum and determine, for example, whether
there is a high-energy cutoff above several PeV and whether there are both galactic and
extra-galactic components in the astrophysical signal.  It will also significantly
increase the sensitivity for the search for cosmogenic neutrinos with much higher
energies.  The PINGU~\cite{Aartsen:2014oha} component of Gen2 will provide increased density of
instrumentation in the current DeepCore portion of IceCube for neutrino oscillation
physics including the mass hierarchy.

Plans also call for a surface array with sufficient detector density to
act as a veto for downward cosmic-ray background.  This would make it possible
to include events generated by charged-current interactions of muon neutrinos
in the ice above the deep detector.  

From the point of view
of cosmic-ray physics, it is important to note that expanding the surface array
in proportion to the area of the deep detector leads to a quadratic increase in
acceptance for coincident events compared to the present detector.  The acceptance for coincident
events of a surface array of area $A_s$ centered above a deep detector of area $A_d$
at depth $d$ is 
\begin{equation}
A\,\Omega\,\approx\,\frac{A_sA_d}{d^2}.
\label{eq:acceptance}
\end{equation}
Taking $d=2$~km, the acceptance of the present IceCube for coincident events
is $\approx 0.25$~km$^2$sr.  With an area of $A_s=A_d\approx 7$~km$^2$
the corresponding number would be a factor of $\approx 50$ larger.
In addition, for purposes of the veto, it is desirable to have a surface array
that extends beyond the footprint of the deep detector.  Studies are
ongoing to optimize the surface component of Gen2 for both veto and cosmic-ray physics.

\section*{Acknowledgments}
I am grateful for helpful discussions with Hermann Kolanoski, Javier Gonzalez
and Hans Dembinski.
The research on which this paper reports is supported in part by the 
U.S. National Science Foundation.  A full list of supporting agencies for the
IceCube Neutrino Observatory may be found at http://icecube.wisc.edu/collaboration/funding.




\bibliographystyle{elsarticle-num}
\bibliography{Gaisser}







\end{document}